\documentclass[aps,prb,fleqn,12pt]{revtex4}
\usepackage{epsfig,color}
\usepackage{graphicx}
\usepackage{pdfpages}
\usepackage{braket}
\usepackage{amssymb,amsmath}
\usepackage{hyperref}
\begin{document}
\title{Magnetic reorientation transition in a three orbital model for {\boldmath $\rm Ca_2 Ru O_4$} --- Interplay of spin-orbit coupling, tetragonal distortion, and Coulomb interactions}
\author{Shubhajyoti Mohapatra and Avinash Singh}
\email{avinas@iitk.ac.in}
\affiliation{Department of Physics, Indian Institute of Technology, Kanpur - 208016, India}
\date{\today} 
\begin{abstract}
Including the orbital off-diagonal spin and charge condensates in the self consistent determination of magnetic order within a realistic three-orbital model for the $4d^4$ compound $\rm Ca_2 Ru O_4$, reveals a host of novel features including strong and anisotropic spin-orbit coupling (SOC) renormalization, coupling of strong orbital magnetic moments to orbital fields, and a magnetic reorientation transition. Highlighting the rich interplay between orbital geometry and overlap, spin-orbit coupling, Coulomb interactions, tetragonal distortion, and staggered octahedral tilting and rotation, our investigation yields a planar antiferromagnetic (AFM) order for moderate tetragonal distortion, with easy $a-b$ plane and easy $b$ axis anisotropies, along with small canting of the dominantly $yz,xz$ orbital moments. With decreasing tetragonal distortion, we find a magnetic reorientation transition from the dominantly planar AFM order to a dominantly $c$ axis ferromagnetic (FM) order with significant $xy$ orbital moment. 
\end{abstract}
\maketitle
\newpage

\section{Introduction}

The interplay of spin-orbit coupling (SOC) with electronic correlations and crystal field splittings has been found to drive various topologically nontrivial phases in condensed matter systems such as topological Mott insulators, quantum spin liquids, and superconducting states.\cite{krempa_AR_2014,Cao_RPP_2018} The $4d$ and $5d$ transition metal oxides containing $\rm Ru^{4+}$, $\rm Os^{4+}$, $\rm Ir^{4+}$, $\rm Ir^{5+}$ ions have emerged as promising candidates exhibiting SOC-induced exotic ground states, magnetic anisotropy effects, and intriguing collective excitations. SOC effects in the $d^5$ systems are more transparent and well understood in terms of the spin-orbital entangled electronic states with nominally filled $J=3/2$ quartet and half-filled magnetically active $J=1/2$ doublets.\cite{bjkim_PRL_2008} The isospin dynamics involving $J$ states provides insight into the experimentally observed magnetic behavior in perovskite iridates as well as iridate heterostructures which are gaining interest as their magnetic properties are much more sensitive to structural distortion compared to pure spin systems due to spin-orbital entanglement.\cite{iridate1_PRB_2017,iridate3_arxiv_2020,iridate4_JMMM_2020,mohapatra_PRB_2019}

However, the situation is very different in $d^4$ systems with four electrons per metal ion. For strong SOC, all four electrons fill the $J=3/2$ sector, leaving the $J=1/2$ sector empty and naturally leading to non-magnetic insulating behavior.\cite{fuchs_PRL_2018} Similarly, for strong Hund's coupling, total spin moment $S = 1$ antiparallel to the orbital moment $L = 1$ leads to total angular momentum $J = 0$ on every metal ion with no magnetism. Thus, both scenarios lead to the non-magnetic $J = 0$ singlet ground state for $d^4$ systems. However, magnetism has been revealed in some double perovskite iridates and ruthenates with $d^4$ electronic configuration, and the origin of magnetism is under investigation.\cite{wang_PRB_2014,cao_PRL_2014,marco_PRB_2015,bhowal_PRB_2015,dey_PRB_2016}

Among $d^4$ systems, the quasi-two-dimensional antiferromagnet $\rm Ca_2RuO_4$ has attracted strong interest. With decreasing temperature, it undergoes a peculiar non-magnetic metal-insulator transition (MIT) at 356 K, and a magnetic transition at $T_{\rm N}\approx$ 113 K with observed magnetic moment of $1.3~ \mu_{\rm B}$. \cite{nakatsuji_JPSP_1997,braden_PRB_1998,alexander_PRB_1999,friedt_PRB_2001} Under high pressure and at low temperature, $\rm Ca_2RuO_4$ undergoes a transition to a ferromagnetic (FM) metallic phase, with maximum $T_{\rm C} \approx 30$ K at 5 GPa pressure,\cite{nakamura_PRB_2002} and the existence of a FM quantum critical point at pressures above 10 GPa is indicated. The MIT is associated with a structural transition from L-phase (long octahedral $c$-axis) to S-phase (short $c$-axis) due to continuous flattening of octahedra till the onset of antiferromagnetic (AFM) order at $T_{\rm N}$.\cite{gorelov_PRL_2010} Compared to the isoelectronic member $\rm Sr_2RuO_4$,\cite{nakatsuji_PRL_2000,friedt_PRB_2001} this system has severe structural distortions due to the small $\rm Ca^{2+}$ size, resulting in compression, rotation, and tilting of the $\rm RuO_6$ octahedra. Thus, the low-temperature phase is characterized by highly distorted $\rm RuO_6$ octahedra and canted AFM order with moments lying along the crystal $b$ axis.\cite{fang_PRB_2001,Kunkemoller_PRL_2015} Such transitions have been identified in temperature,\cite{nakatsuji_PRB_2000} hydrostatic pressure,\cite{steffens_PRB_2005} epitaxial strain,\cite{dietl_APL_2018} chemical substitution,\cite{nakatsuji_PRL_2000,nakamura_PRB_2002,steffens_PRB_2011} and electrical current\cite{nakamura_SREP_2013,okazaki2_JPSJ_2013} studies of $\rm Ca_2RuO_4$. 

In the isoelectronic series $\rm Ca_{2-x}Sr_x RuO_4$, the ground state has been successively driven from the AFM insulator ($x<0.2$) to an AFM correlated metal ($0.2<x<0.5$), a nearly FM metal ($x\sim 0.5$), and finally to a non-magnetic two-dimensional Fermi liquid ($x\sim 2$). Since the substitution is isovalent, the dominant effects are structural modifications due to larger Sr ionic size.\cite{nakatsuji_PRB_2000} With increasing $x$, the distortion occurs in steps, resulting in removal of first the flattening of the octahedra, then the tilting, and finally the rotation around the $c$ axis.\cite{friedt_PRB_2001,fang_PRB_2001,fang_PRB_2004} Although the substitution is isovalent, the magnetism of $\rm Ca_{2-x}Sr_{x}RuO_4$ is affected in the sequence given above by the changes in orbital hybridization resulting from substitution induced structural distortions. 


In the literature, mainly two different scenarios have been discussed for classifying the magnetism in $\rm Ca_2RuO_4$. In the first, octahedral compression induced large tetragonal crystal field ($\approx 0.3$ eV) lifts the degeneracy of the $t_{\rm 2g}$ orbitals by lowering the $xy$ orbital energy. Based on DFT calculations,\cite{fang_PRB_2004,liebsch_PRL_2007,gorelov_PRL_2010,zhang_PRB_2017} which agree with X-ray scattering as well as angle resolved photoemission spectroscopy (ARPES) studies,\cite{fatuzzo_PRB_2015,sutter_NATCOM_2017} the $xy$ orbital is nominally filled, and the half-filled $yz,xz$ orbitals form a spin $S=1$ state. Further, low octahedral symmetry around the Ru ion is believed to quench the orbital moment completely. Thus, the ordering of $S=1$ spins supports a more conventional explanation for the magnetism with a negligible role of SOC. However, the presence of the strong in-plane anisotropy in the magnon dispersion indicates the importance of SOC in tuning the magnetic anisotropy in the system.\cite{Kunkemoller_PRL_2015}

\begin{figure}
\vspace*{0mm}
\hspace*{0mm}
\psfig{figure=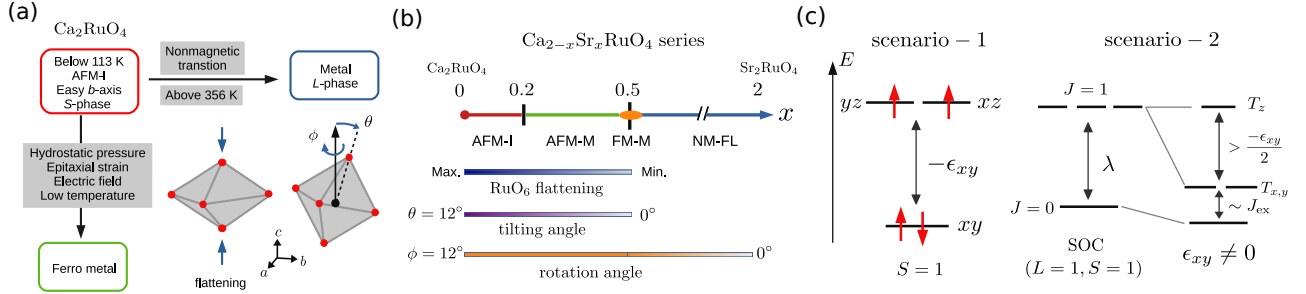,angle=0,width=170mm,angle=0}
\caption{Summary of the physical properties of (a) layered $\rm Cu_2RuO_4$ showing transitions from AFM insulator (AFM-I) to non-magnetic metal or ferromagnetic metal (FM-M) induced by different agents, and (b) isoelectronic series ${\rm Ca}_{2-x} {\rm Sr}_x{\rm RuO_4}$ showing successive transition from AFM-I to non-magnetic Fermi liquid (NM-FL), through AFM-M and FM-M states. With increasing $x$, the distortions occur in steps. (c) Tetragonal field $\epsilon_{xy}$ and Hund's coupling stabilized $L$=0, $S$=1 state in absence of SOC (scenario-1). Strong SOC picture (scenario-2) showing nonmagnetic $J$=0 ground state and $J$=1 triplet excited state, which further splits into singlet ($T_z$) and degenerate doublet ($T_{x,y}$) in presence of $\epsilon_{xy}$. The energy difference $[E(T_{x,y})-E(J$=0$)]$ comparable to the exchange energy ($J_{\rm ex}$) induces magnetic ordering.} 
\label{summary}
\end{figure}

In the second scenario, $\rm Ca_2RuO_4$, with only moderate SOC strength, has been argued as a possible candidate for excitonic antiferromagnetism. If the superexchange involving excited magnetic states (triplet $J = 1$) is strong enough to compete with the singlet-triplet splitting caused by SOC, the on-site wave function becomes a superposition of $J = 0,1$ states and acquires a magnetic moment.\cite{khaliullin_PRL_2013,akbari_PRB_2014,khomskii_2014,meetei_PRB_2015,feldmaier_arxiv_2019} This picture is supported by the observed unconventional magnetic excitation spectra from the recent inelastic neutron scattering (INS) and resonant inelasic X-ray scattering (RIXS) experiments.\cite{jain_NATPHY_2017,fatuzzo_PRB_2015,das_PRX_2018,gretarsson_PRB_2019} Spin-wave dispersion in the INS study has revealed a global maximum at the Brillouin zone center, which is in sharp contrast to the $S=1$ quantum Heisenberg antiferromagnet (QHAF), and has been interpreted as a sign of such excitonic magnetism in $\rm Ca_2RuO_4$.\cite{jain_NATPHY_2017} The various properties of this system and theoretical scenarios as discussed above are summarized in Fig. \ref{summary}.

While numerous computational and experimental techniques have been applied,\cite{puchkov_PRL_1998,park_JPCM_2001,liu_PRB_2011,sutter_NATCOM_2017,acharya_JPCOM_2018,bertinshaw_PRL_2019,hao_arxiv_2019} very little is known about the electronic band structure of $\rm Ca_2RuO_4$ in the low-temperature AFM state. Earlier numerical calculations within three orbital models have adopted simplified Hamiltonians to discuss the mechanism of metal-insulator transition and magnetism.\cite{kaushal_PRB_2017,sato_PRB_2019} However, realistic hoppings, structural distortions, SOC, and electronic correlations were not considered on an equal footing in these simplistic models. Earlier works have also lacked in fully accounting for the Coulomb interaction effects, especially those associated with orbital off-diagonal spin and charge correlations. Indeed, the effective SOC strength $\sim 200$ meV extracted from ARPES and RIXS studies\cite{mizokawa_PRL_2001,fatuzzo_PRB_2015} indicates a strong correlation-induced enhancement compared to the predicted theoretical value $\sim 100$ meV.\cite{khaliullin_PRL_2013,akbari_PRB_2014}

The richness and complexity displayed in structural, magnetic, and transport properties of this system, along with intimate couplings between lattice, spin, and charge degrees of freedom, have led to difficulty in realistic modeling of these phenomena. Classification of the nature of magnetic ground state and the role of SOC and distortion effects in tuning the magnetic behavior of $\rm Ca_2RuO_4$ therefore remains far from being well understood. A delicate interplay of different Coulomb interaction terms with SOC may lead to complex and nontrivial behavior of orbital and spin degrees of freedom. Investigation of magnetic ordering, anisotropy, and  electronic band structure in $\rm Ca_2RuO_4$ by incorporating the SOC, structural distortions, and multi-orbital Coulomb interaction terms on an equal footing is therefore of strong interest.  

For a multi-orbital interacting electron system, a general treatment of the various Coulomb interaction terms in the Hartree-Fock (HF) approximation yields, besides the contributions from the normal (orbital diagonal) spin and charge density condensates, additional contributions involving orbital off-diagonal condensates. Since the SOC and orbital angular momentum terms involve orbital off-diagonal one-body operators, due to interplay between strong SOC-induced spin-orbital correlations and Coulomb interactions, $\rm Ca_2RuO_4$ presents a case where the off-diagonal condensates should play an important role in determining the magnetic order and anisotropy. However, these aspects have not been systematically investigated within the itinerant electron picture. 

In this work, all orbital off-diagonal spin $\langle \psi_\mu^\dagger \makebox{\boldmath $\sigma$} \psi_\nu \rangle$ and charge $\langle \psi_\mu^\dagger {\bf 1} \psi_\nu \rangle$ condensates will therefore be included, and a self consistent determination of magnetic order and anisotropy will be carried out within a realistic three-orbital interacting electron model for $\rm Ca_2RuO_4$ in the $t_{\rm 2g}$ manifold of the $\mu,\nu=yz,xz,xy$ orbitals. The orbital off-diagonal spin and charge condensates will be seen to result in strong and anisotropic SOC renormalization and strong orbital magnetic moments $\langle L_{x,y,z}\rangle$ in the magnetic ground state. We will first focus on the planar AFM order with dominantly $yz,xz$ moments, which is realized for moderate tetragonal distortion. However, with decreasing tetragonal distortion, we find a magnetic reorientation transition to a dominantly $c$ axis ferromagnetic (FM) order, as seen in high-pressure investigations of $\rm Ca_2RuO_4$.\cite{nakamura_PRB_2002}  

The structure of this paper is as follows. After introducing the three-orbital model and Coulomb interaction terms in Sec. II, the SOC-induced easy-plane anisotropy and the octahedral tilting induced easy-axis anisotropy are discussed in Secs. III and IV. Results of the self-consistent determination of magnetic order including all orbital off-diagonal spin and charge condensates in the HF approximation are presented in Sec. V, together with the orbital resolved electronic band structure. The orbital magnetic moments and Coulomb interaction induced anisotropic SOC renormalization are discussed in Sec. VI, and the magnetic reorientation transition in Sec. VII. After some observations on the strongly coupled spin-orbital fluctuations in Sec. VIII, conclusions are finally presented in Sec. IX. 

\section{Three orbital model and Coulomb interactions}
\begin{figure}
\vspace*{0mm}
\hspace*{0mm}
\psfig{figure=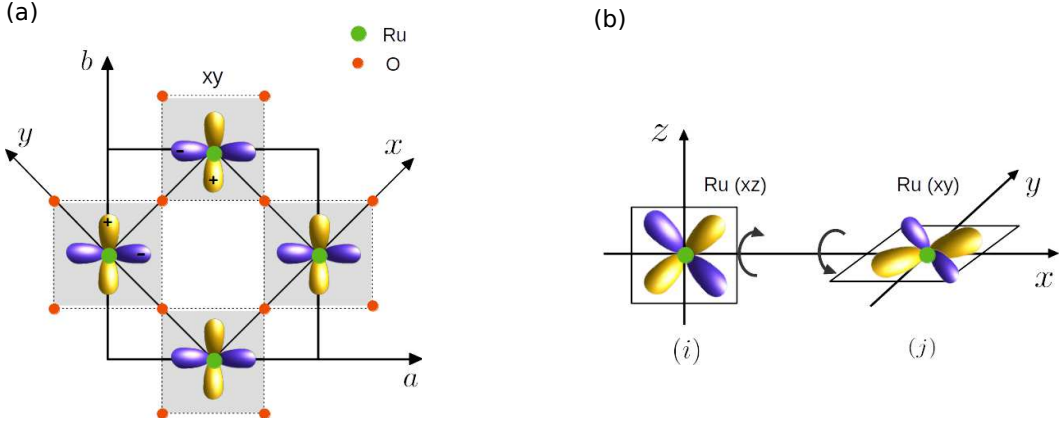,angle=0,width=140mm,angle=0}
\caption{(a) The common spin-orbital coordinate axes ($x-y$) along the Ru-O-Ru directions, shown along with the crystal axes $a,b$. (b) Octahedral tilting about the crystal $a$ axis is resolved along the $x,y$ axes, resulting in orbital mixing hopping terms between the $xy$ and $yz,xz$ orbitals.} 
\label{axes}
\end{figure}

In the three-orbital ($\mu=yz,xz,xy$), two-spin ($\sigma=\uparrow,\downarrow$) basis defined with respect to a common spin-orbital coordinate axes (Fig. \ref{axes}), we consider the Hamiltonian ${\cal H} = {\cal H}_{\rm SOC} + {\cal H}_{\rm cf} + {\cal H}_{\rm band} + {\cal H}_{\rm int}$ within the $t_{\rm 2g}$ manifold. The spin-orbit coupling term ${\cal H}_{\rm SOC}$, which explicitly breaks SU(2) spin rotation symmetry and therefore generates anisotropic magnetic interactions from its interplay with other Hamiltonian terms, will be introduced in the next section.

For the band and crystal field terms together, we consider:
\begin{eqnarray}
{\cal H}_{\rm band+cf} &=& 
\sum_{{\bf k} \sigma s} \psi_{{\bf k} \sigma s}^{\dagger} \left [ \begin{pmatrix}
{\epsilon_{\bf k} ^{yz}}^\prime & 0 & 0 \\
0 & {\epsilon_{\bf k} ^{xz}}^\prime & 0 \\
0 & 0 & {\epsilon_{\bf k} ^{xy}}^\prime + \epsilon_{xy}\end{pmatrix} \delta_{s s^\prime}
+ \begin{pmatrix}
\epsilon_{\bf k}^{yz} & \epsilon_{\bf k}^{yz|xz} & \epsilon_{\bf k}^{yz|xy} \\
-\epsilon_{\bf k} ^{yz|xz} & \epsilon_{\bf k} ^{xz} & \epsilon_{\bf k}^{xz|xy} \\
-\epsilon_{\bf k}^{yz|xy} & -\epsilon_{\bf k}^{xz|xy} & \epsilon_{\bf k} ^{xy} \end{pmatrix} \delta_{\bar{s} s^\prime } \right] \psi_{{\bf k} \sigma s^\prime} \nonumber \\
\label{three_orb_two_sub}
\end{eqnarray} 
in the composite three-orbital, two-sublattice ($s,s'={\rm A,B}$) basis. Here the energy offset $\epsilon_{xy}$ (relative to the degenerate $yz/xz$ orbitals) represents the tetragonal distortion induced crystal field effect, and the band dispersion terms in the two groups, corresponding to hopping terms connecting the same and opposite sublattice(s), are given by: 
\begin{eqnarray}
\epsilon_{\bf k} ^{xy} &=& -2t_1(\cos{k_x} + \cos{k_y}) \nonumber \\
{\epsilon_{\bf k} ^{xy}} ^{\prime} &=& - 4t_2\cos{k_x}\cos{k_y} - \> 2t_3(\cos{2{k_x}} + \cos{2{k_y}}) \nonumber \\
\epsilon_{\bf k} ^{yz} &=& -2t_5\cos{k_x} -2t_4 \cos{k_y} \nonumber \\
\epsilon_{\bf k} ^{xz} &=& -2t_4\cos{k_x} -2t_5 \cos{k_y}  \nonumber \\
\epsilon_{\bf k} ^{yz|xz} &=&  -2t_{m1}(\cos{k_x} + \cos{k_y}) \nonumber \\
\epsilon_{\bf k} ^{xz|xy} &=&  -2t_{m2}(2\cos{k_x} + \cos{k_y}) \nonumber \\
\epsilon_{\bf k} ^{yz|xy} &=&  -2t_{m3}(\cos{k_x} + 2\cos{k_y}) . 
\label{band}
\end{eqnarray}

Here $t_1$, $t_2$, $t_3$ are respectively the first, second, and third neighbor hopping terms for the $xy$ orbital. For the $yz$ ($xz$) orbital, $t_4$ and $t_5$ are the NN hopping terms in $y$ $(x)$ and $x$ $(y)$ directions, respectively, corresponding to $\pi$ and $\delta$ orbital overlaps. Octahedral rotation and tilting induced orbital mixings are represented by the NN hopping terms $t_{m1}$ (between $yz$ and $xz$) and $t_{m2},t_{m3}$ (between $xy$ and $xz,yz$). We have taken hopping parameter values: ($t_1$, $t_2$, $t_3$, $t_4$, $t_5$)=$(-1.0, 0.5, 0, -1.0, 0.2)$, and for the orbital mixing terms: $t_{m1}$=0.2 and $t_{m2}$=$t_{m3}$=0.15 ($\approx 0.2/\sqrt{2}$), all in units of the realistic hopping energy scale $|t_1|$=200meV.\cite{khaliullin_PRL_2013,akbari_PRB_2014,feldmaier_arxiv_2019} 
The choice $t_{m2}=t_{m3}$ corresponds to the octahedral tilting axis oriented along the $\pm(-\hat{x}+\hat{y})$ direction, which is equivalent to the crystal $\mp a$ direction (Fig. \ref{axes}). The $t_{m1}$ and $t_{m2,m3}$ values taken above approximately correspond to octahedral rotation and tilting angles of about $12^\circ$ ($\approx 0.2$ rad) as reported in experimental studies.\cite{steffens_PRB_2005} 

For the on-site Coulomb interaction terms in the $t_{2g}$ basis ($\mu,\nu=yz,xz,xy$), we consider:
\begin{eqnarray}
{\cal H}_{\rm int} &=& U\sum_{i,\mu}{n_{i\mu\uparrow}n_{i\mu\downarrow}} + U^\prime \sum_{i,\mu < \nu,\sigma} {n_{i\mu\sigma} n_{i\nu\overline{\sigma}}} + (U^\prime - J_{\mathrm H}) \sum_{i,\mu < \nu,\sigma}{n_{i\mu\sigma}n_{i\nu\sigma}} \nonumber\\ 
&+& J_{\mathrm H} \sum_{i,\mu \ne \nu} {a_{i \mu \uparrow}^{\dagger}a_{i \nu\downarrow}^{\dagger}a_{i \mu \downarrow} a_{i \nu \uparrow}} + J_{\mathrm P} \sum_{i,\mu \ne \nu} {a_{i \mu \uparrow}^{\dagger} a_{i \mu\downarrow}^{\dagger}a_{i \nu \downarrow} a_{i \nu \uparrow}} \nonumber\\ 
&=& U\sum_{i,\mu}{n_{i\mu\uparrow}n_{i\mu\downarrow}} + U^{\prime \prime}\sum_{i,\mu<\nu} n_{i\mu} n_{i\nu} - 2J_{\mathrm H} \sum_{i,\mu<\nu} {\bf S}_{i\mu}.{\bf S}_{i\nu} 
+J_{\mathrm P} \sum_{i,\mu \ne \nu} a_{i \mu \uparrow}^{\dagger} a_{i \mu\downarrow}^{\dagger}a_{i \nu \downarrow} a_{i \nu \uparrow} 
\label{h_int}
\end{eqnarray} 
including the intra-orbital $(U)$ and inter-orbital $(U')$ density interaction terms, the Hund's coupling term $(J_{\rm H})$, and the pair hopping interaction term $(J_{\rm P})$, with $U^{\prime\prime} \equiv U^\prime-J_{\rm H}/2=U-5J_{\rm H}/2$ from the spherical symmetry condition $U^\prime=U-2J_{\mathrm H}$. Here $a_{i\mu\sigma}^{\dagger}$ and $a_{i\mu \sigma}$ are the electron creation and annihilation operators for site $i$, orbital $\mu$, spin $\sigma=\uparrow ,\downarrow$, and the density operator $n_{i\mu\sigma}=a_{i\mu\sigma}^\dagger a_{i\mu\sigma}$, total density operator $n_{i\mu}=n_{i\mu\uparrow}+n_{i\mu\downarrow}=\psi_{i\mu}^\dagger \psi_{i\mu}$, and spin density operator ${\bf S}_{i\mu} = \psi_{i\mu}^\dagger ${\boldmath $\sigma$}$ \psi_{i\mu}$, where $\psi_{i\mu}^\dagger=(a_{i\mu\uparrow}^{\dagger} \; a_{i\mu\downarrow}^{\dagger})$. All interaction terms above are SU(2) invariant and thus possess spin rotation symmetry in real-spin space. In the following, we will take $U=8$ in the energy scale unit (200 meV) and $J_{\rm H}=U/5$, so that $U=1.6$eV, $U^{\prime\prime}=U/2=0.8$eV, and $J_{\rm H}=0.32$eV. These are comparable to reported values extracted from RIXS ($J_{\rm H}=0.34$eV) and ARPES ($J_{\rm H}=0.4$eV) studies.\cite{gretarsson_PRB_2019,sutter_NATCOM_2017} 

For moderate tetragonal distortion ($\epsilon_{xy}\approx -1$), the $xy$ orbital in the $4d^4$ compound $\rm Ca_2 Ru O_4$ is nominally doubly occupied and magnetically inactive, while the nominally half-filled and magnetically active $yz,xz$ orbitals yield an effectively two-orbital magnetic system. Hund's coupling between the two $S=1/2$ spins results in low-lying (in-phase) and appreciably gapped (out-of-phase) spin fluctuation modes. The in-phase modes of the $yz,xz$ orbital $S=1/2$ spins correspond to an effective $S=1$ spin system. However, the rich interplay between SOC, Coulomb interaction, octahedral rotations, and tetragonal distortion results in complex magnetic behaviour which crucially involves the $xy$ orbital and is therefore beyond the above simplistic picture. Before proceeding with the self-consistent determination of magnetic order (Sec. V), some of the important physical elements are individually discussed below. 

\section{SOC induced easy plane anisotropy}
The bare spin-orbit coupling term (for site $i$) can be written in spin space as:
\begin{eqnarray} 
H_{\rm SOC} (i) & = & -\lambda {\bf L}.{\bf S} = -\lambda (L_z S_z + L_x S_x + L_y S_y) \nonumber \\ 
&=& \left [ \begin{pmatrix} \psi_{yz \uparrow}^\dagger & \psi_{yz \downarrow}^\dagger \end{pmatrix}
\begin{pmatrix} i \sigma_z \lambda /2 \end{pmatrix} 
\begin{pmatrix} \psi_{xz \uparrow} \\ \psi_{xz \downarrow} \end{pmatrix}
+ \begin{pmatrix} \psi_{xz \uparrow}^\dagger & \psi_{xz \downarrow}^\dagger \end{pmatrix}
\begin{pmatrix} i \sigma_x \lambda /2 \end{pmatrix} 
\begin{pmatrix} \psi_{xy \uparrow} \\ \psi_{xy \downarrow} \end{pmatrix} \right . \nonumber \\
& + & \left . \begin{pmatrix} \psi_{xy \uparrow}^\dagger & \psi_{xy \downarrow}^\dagger \end{pmatrix}
\begin{pmatrix} i \sigma_y \lambda /2 \end{pmatrix} 
\begin{pmatrix} \psi_{yz \uparrow} \\ \psi_{yz \downarrow} \end{pmatrix} \right ] + {\rm H.c.}
\label{soc}
\end{eqnarray}
which explicitly shows the SU(2) spin rotation symmetry breaking. Here we have used the matrix representations:
\begin{equation}
L_z = \begin{pmatrix} 0&-i&0 \\ i&0&0 \\ 0&0&0 \end{pmatrix} ,\;\;
L_x = \begin{pmatrix} 0&0&0 \\ 0&0&-i \\ 0&i&0 \end{pmatrix} ,\;\;
L_y = \begin{pmatrix} 0&0&i \\ 0&0&0 \\ -i&0&0 \end{pmatrix} ,\;\;
\end{equation}
for the orbital angular momentum operators in the three-orbital $(yz,xz,xy)$ basis. 

As the orbital ``hopping" terms in Eq. (\ref{soc}) have the same form as spin-dependent hopping terms $i${\boldmath $\sigma . t'_{ij}$}, carrying out the strong-coupling expansion\cite{hc_JMMM_2019} for the $-\lambda L_z S_z$ term to second order in $\lambda$ yields the anisotropic diagonal (AD) intra-site interactions:
\begin{equation}
[H^{(2)}_{\rm eff}]_{\rm AD}^{(z)}(i) = \frac{4 (\lambda/2)^2 }{U} \left [ S_{yz}^z S_{xz}^z - (S_{yz}^x S_{xz}^x  + S_{yz}^y S_{xz}^y) \right ] 
\label{h_eff}
\end{equation}
between $yz,xz$ moments in these nominally half-filled orbitals. Corresponding to an effective single-ion anisotropy (SIA), this term explicitly yields preferential $x-y$ plane ordering for parallel $yz,xz$ moments, as enforced by the relatively stronger Hund's coupling. 

For later reference, we note here that condensates of the orbital off-diagonal one-body operators as in Eq. (\ref{soc}) directly yield physical quantities such as orbital magnetic moments and spin-orbital correlations:
\begin{eqnarray}
\langle L_\alpha \rangle & = & -i \left [ \langle \psi_\mu^\dagger \psi_\nu\rangle - \langle \psi_\mu^\dagger \psi_\nu\rangle^* \right ] = 2\ {\rm Im}\langle \psi_\mu^\dagger \psi_\nu\rangle \nonumber \\
\langle L_\alpha S_\alpha \rangle & = & -i \left [ \langle \psi_\mu^\dagger \sigma_\alpha \psi_\nu\rangle - \langle \psi_\mu^\dagger \sigma_\alpha \psi_\nu\rangle^* \right]/2 = {\rm Im}\langle \psi_\mu^\dagger \sigma_\alpha \psi_\nu\rangle\nonumber \\
\lambda^{\rm int}_\alpha & \approx & U'' \langle L_\alpha S_\alpha \rangle 
\label{phys_quan}
\end{eqnarray}
where the orbital pair ($\mu,\nu$) corresponds to the component $\alpha=x,y,z$, and the last yields the interaction induced SOC renormalization, as discussed in Sec. VI.  

\section{Octahedral tilting and easy-axis anisotropy}
While SOC directly induces an easy $x-y$ plane anisotropy, interplay between the staggered octahedral tilting in $\rm Ca_2RuO_4$ and SOC yields an easy-axis anisotropy along the $\hat{x}+\hat{y}$ direction, which is same as the crystal $b$ direction. Octahedral tilting generates orbital mixing hopping terms between $xy$ and $yz,xz$ orbitals (Eq. \ref{band}). These normal NN hopping terms, together with the local spin-flip SOC mixing terms between $xy$ and $yz,xz$ orbitals, lead to effective spin-dependent NN hopping terms:
\begin{equation}
{\cal H}_{\rm eff} ' = \sum_{\langle i,j \rangle,\mu} \psi_{i\mu} ^\dagger [-i\makebox{\boldmath $\sigma$}.{\bf t'}] \psi_{j\mu} + {\rm H.c.} 
\label{sdhterms}
\end{equation}
for the magnetically active ($\mu=yz,xz$) orbitals. The hopping terms are bond dependent, with only finite $t_x'$ ($t_y'$) between $xz$ ($yz$) orbital in the $x$ ($y$) direction. Within the usual strong-coupling expansion, the combination of the normal ($t$) and spin-dependent ($t_x',t_y'$) hopping terms generates Dzyaloshinski-Moriya (DM) interaction terms in the effective spin model:
\begin{eqnarray}
[H_{\rm eff}^{(2)}]_{\rm DM}^{(x,y)} &=& \frac{8tt_x'}{U} \sum_{\langle i,j \rangle_x} \hat{x}.({\bf S}_{i,xz} \times {\bf S}_{j,xz}) 
+ \frac{8tt_y'}{U} \sum_{\langle i,j\rangle_y} \hat{y}.({\bf S}_{i,yz} \times {\bf S}_{j,yz}) \nonumber \\ 
& \approx & \frac{8t|t_x'|}{U} \sum_{\langle i,j \rangle} 
(-\hat{x} + \hat{y}).({\bf S}_i \times {\bf S}_j) 
\end{eqnarray}
for $t_x'=-t_y'=-$ive and ${\bf S}_{i,xz} \approx {\bf S}_{i,yz}$ due to the relatively much stronger Hund's coupling. The effective DM axis ($-\hat{x} + \hat{y}$) is along the octahedral tilting axis, which is same as the crystal $-a$ axis (Fig. \ref{axes}).  

\begin{figure}
\vspace*{0mm}
\hspace*{0mm}
\psfig{figure=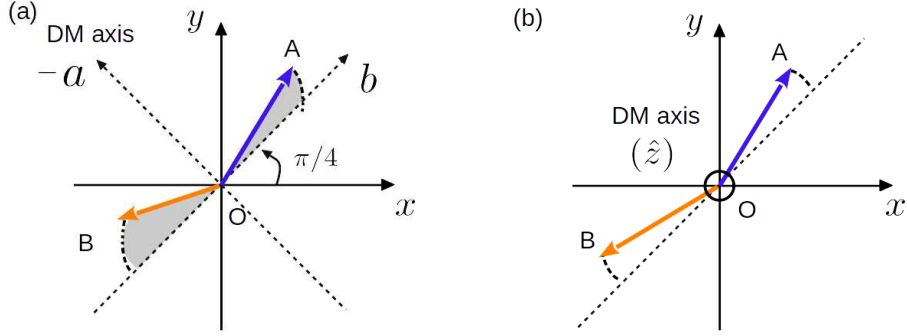,angle=0,width=120mm,angle=0}
\caption{Spin cantings about the (a) crystal $a$ axis and (b) crystal $c$ axis, due to the effective DM interactions induced by the staggered octahedral tilting and rotation, respectively. Octahedral tilting about crystal $a$ axis yields the perpendicular (crystal $b$)  direction as the magnetic easy axis.} 
\label{canting}
\end{figure}

The easy-axis anisotropy as well as spin canting in the $z$ direction follow directly from the above DM interaction, which induces spin canting about the DM axis and favors spins lying in the perpendicular plane. Intersection of the perpendicular plane $(\phi=\pi/4,z)$ and the SOC-induced easy $x-y$ plane yields $\phi=\pi/4$ as the easy-axis direction, and canting about the DM axis yields spin canting in the $z$ direction, as shown in Fig. \ref{canting}(a). 

In close analogy with the above effects of octahedral tilting, the staggered octahedral rotation about the crystal $c$ axis leads to orbital mixing hopping terms between $yz,xz$ orbitals on NN sites, and hence to effective spin-dependent NN hopping terms $t_z'$ in Eq. (\ref{sdhterms}). The resulting effective DM term $-(8tt_z'/U)\hat{z}.({\bf S}_i \times {\bf S}_j)$ causes spin canting about the crystal $c$ axis, as shown in Fig. \ref{canting}(b). The easy-axis anisotropy as well as the two spin cantings of the dominant $yz,xz$ moments are confirmed in the full self-consistent calculation discussed below. Also, the effective spin dependent hopping terms discussed above are explicitly confirmed from the electronic band structure features in the self consistent state.

Before continuing with the other important physical elements, it is convenient to first systematically introduce the different Coulomb interaction contributions in the HF theory. Contributions involving the orbital off-diagonal spin and charge condensates naturally lead to interaction induced SOC renormalization and coupling of orbital magnetic moments to orbital fields.  

\section{Self-consistent determination of magnetic order}
We consider the various Coulomb interaction terms in Eq. (\ref{h_int}) in the HF approximation, focussing first on the terms with normal (orbital diagonal) spin and charge condensates. The resulting local spin and charge terms can be written as:
\begin{equation}
[{\cal H}_{\rm int}^{\rm HF}]_{\rm normal} = \sum_{i\mu} \psi_{i\mu}^{\dagger} \left [
-\makebox{\boldmath $\sigma . \Delta$}_{i\mu} + {\cal E}_{i\mu} {\bf 1} \right ] \psi_{i\mu} 
\label{h_hf} 
\end{equation}  
where the spin and charge fields are self-consistently determined from:
\begin{eqnarray}
2\Delta_{i\mu}^\alpha &=& U\langle \sigma_{i\mu}^\alpha \rangle + J_{\rm H} \sum_{\nu < \mu} \langle \sigma_{i\nu}^\alpha \rangle \;\;\;\;\;(\alpha=x,y,z) \nonumber \\
{\cal E}_{i\mu} &=& \frac{U\langle n_{i\mu}\rangle}{2} + U'' \sum_{\nu < \mu} \langle n_{i\nu} \rangle 
\label{selfcon}
\end{eqnarray}
in terms of the local charge density $\langle n_{i\mu}\rangle$ and the spin density components $\langle \sigma_{i\mu}^\alpha \rangle$. For $\langle n_{yz}\rangle=\langle n_{xz}\rangle$, the Coulomb renormalized tetragonal splitting is obtained as:
\begin{eqnarray}
\tilde{\delta}_{\rm tet} &=& \tilde{\epsilon}_{xz,yz} - \tilde{\epsilon}_{xy} = (\epsilon_{xz,yz} - \epsilon_{xy}) 
+ \left [{\cal E}_{yz,xz} - {\cal E}_{xy} \right ] \nonumber \\
& = & \delta_{\rm tet} + \left [\frac{U\langle n_{yz,xz}\rangle}{2} + U'' \langle n_{yz,xz} + n_{xy}\rangle \right ] - \left [\frac{U\langle n_{xy}\rangle}{2} + 2U'' \langle n_{yz,xz}\rangle \right ] \nonumber \\
& = & \delta_{\rm tet} + (U'' -U/2)\langle n_{xy} - n_{yz,xz}\rangle 
\end{eqnarray}
which shows that the Coulomb renormalization identically vanishes for the realistic relationship $U''=U/2$ for $4d$ orbitals, as discussed in Sec. II. 

There are additional contributions in the HF approximation resulting from orbital off-diagonal spin and charge condensates which are finite due to the SOC induced spin-orbital correlations. The contributions corresponding to different Coulomb interaction terms are summarized in the Appendix, and can be grouped in analogy with Eq. (\ref{h_hf}) as:
\begin{equation}
[{\cal H}_{\rm int}^{\rm HF}]_{\rm OOD} = \sum_{i,\mu < \nu} \psi_{i\mu}^{\dagger} \left [
-\makebox{\boldmath $\sigma . \Delta$}_{i\mu\nu} + {\cal E}_{i\mu\nu} {\bf 1} \right ] \psi_{i\nu} 
\label{h_hf_od} 
\end{equation}  
where the orbital off-diagonal spin and charge fields are self-consistently determined from:
\begin{eqnarray}
\makebox{\boldmath $\Delta$}_{i\mu\nu} &=& \left (\frac{U''}{2} + \frac{J_{\rm H}}{4} \right ) \langle \makebox{\boldmath $\sigma$}_{i\nu\mu} \rangle + \left (\frac{J_{\rm P}}{2} \right ) \langle \makebox{\boldmath $\sigma$}_{i\mu\nu} \rangle \nonumber \\
{\cal E}_{i\mu\nu} &=& \left (-\frac{U''}{2} + \frac{3J_{\rm H}}{4} \right ) \langle n_{i\nu\mu} \rangle + \left (\frac{J_{\rm P}}{2}\right ) \langle n_{i\mu\nu} \rangle  
\label{sc_od}
\end{eqnarray}
in terms of the corresponding condensates $\langle \makebox{\boldmath $\sigma$}_{i\nu\mu}\rangle \equiv \langle \psi_{i\nu}^{\dagger} \makebox{\boldmath $\sigma$} \psi_{i\mu} \rangle$ and $\langle n_{i\nu\mu} \rangle \equiv \langle \psi_{i\nu}^{\dagger} {\bf 1} \psi_{i\mu} \rangle$. The spin and charge condensates in Eqs. \ref{selfcon} and \ref{sc_od} are evaluated using the eigenfunctions ($\phi_{\bf k}$) and eigenvalues ($E_{\bf k}$) of the full Hamiltonian in the given basis including the interaction contributions $[{\cal H}_{\rm int}^{\rm HF}]$ (Eqs. \ref{h_hf} and \ref{h_hf_od}) using:
\begin{equation}
\langle \sigma^\alpha _{i\mu\nu} \rangle \equiv 
\langle \psi_{i\mu} ^\dagger \sigma^\alpha \psi_{i\nu} \rangle =
\sum_{\bf k}^{E_{\bf k} < E_{\rm F} } (\phi_{{\bf k}\mu s \uparrow}^* \;  \phi_{{\bf k}\mu s \downarrow}^* ) [\sigma^\alpha] \left ( \begin{array}{c} \phi_{{\bf k}\nu s\uparrow} \\ \phi_{{\bf k}\nu s\downarrow} \end{array} \right ) 
\end{equation}
for site $i$ on the $s=A/B$ sublattice, and similarly for the charge condensates $\langle n_{i\mu\nu} \rangle \equiv \langle \psi_{i\mu}^{\dagger} {\bf 1} \psi_{i\nu} \rangle$, with the Pauli matrices $[\sigma^\alpha]$ replaced by the unit matrix $[{\bf 1}]$. The normal spin and charge condensates correspond to $\nu=\mu$.

Results of the full self consistent calculation including all spin and charge condensates (orbital diagonal and off-diagonal) are presented below. For each orbital pair ($\mu,\nu$) = ($yz,xz$), ($xz,xy$), ($xy,yz$), there are three components ($\alpha=x,y,z$) for the spin condensates $\langle \psi_\mu^\dagger \sigma_\alpha \psi_\nu \rangle$ and one charge condensate $\langle \psi_\mu^\dagger {\bf 1} \psi_\nu \rangle$. This is analogous to the three-plus-one normal spin and charge condensates for each of the three orbitals $\mu=yz,xz,xy$. The magnetization and density values for the three orbitals are presented in Table I, all off-diagonal spin and charge condensates in Table II, and the renormalized SOC values and orbital magnetic moments in Table III. Here $U=8$, $\epsilon_{xy}=-0.8$, the bare SOC strength $\lambda^{\rm bare}=1$, and the staggered octahedral rotation ($t_{m1}=0.2$) and tilting ($t_{m2}=t_{m3}=0.15$) have been included. 

\begin{table} \vspace*{-10mm}
\caption{Self consistently determined magnetization and density values for the three orbitals ($\mu$) on the two sublattices ($s$).} 
\centering 
\begin{tabular}{l r r r r} \\  
\hline\hline 
$\mu$ (s) & $m_\mu^x$ & $m_\mu^y$ & $m_\mu^z$ & $n_\mu$ \\ [0.5ex]
\hline 
$yz$ (A) & 0.472 & 0.578 & 0.153 & 1.177  \\[1ex]
$xz$ (A) & 0.459 & 0.647 & 0.163 & 1.133 \\[1ex]
$xy$ (A) & 0.113 & 0.179 & 0.101 & 1.690 \\[1ex] \hline 
$yz$ (B) & $-$0.647 & $-$0.459 & 0.163 & 1.133 \\[1ex]
$xz$ (B) & $-$0.578 & $-$0.472 & 0.153 & 1.177 \\[1ex]
$xy$ (B) & $-$0.179 & $-$0.113 & 0.101 & 1.690 \\[1ex] \hline 
\end{tabular}
\label{table1}
\end{table}

As seen from Table I, the dominant $yz,xz$ moments show the expected cantings in and about the $z$ direction due to the octahedral tilting and rotation (Sec. IV). However, there is an additional small relative canting between the $yz,xz$ moments. To understand the origin of this effect, we consider the real part of the off-diagonal charge condensate $\langle \psi_{xz}^\dagger \psi_{yz} \rangle$ as given in Table II. The corresponding charge term in Eq. (\ref{h_hf_od}) yields a normal ``hopping" term $-(\lambda_0/2) \psi_{yz}^\dagger \psi_{xz}$, and the combination of this normal and spin-dependent $\psi_{yz}^\dagger (i\sigma_z \lambda_z/2) \psi_{xz}$ ``hopping" terms yields an effective intra-site DM interaction:
\begin{equation}
[H_{\rm eff}^{(2)}]_{\rm DM}^{(z)}(i) = -\frac{8(\lambda_0/2)(\lambda_z/2)}{U} \hat{z}.\left ({\bf S}_{yz} \times {\bf S}_{xz} \right ) 
\end{equation}
which leads to relative canting between the $yz$ and $xz$ moments about the $z$ axis. The overall $-$ive sign of the DM term favors canting of ${\bf S}_{yz}$ towards $x$ axis and ${\bf S}_{xz}$ towards $y$ axis. Repeating the calculation with the same parameters as above but without the octahedral rotation, so that the overall canting about the $z$ direction is suppressed, yields magnetization values $m_{yz}^x=m_{xz}^y=\pm 0.56$ and $m_{yz}^y= m_{xz}^x=\pm 0.52$ on A and B sublattices, which clearly show this relative canting effect.   

\begin{figure}[t]
\vspace*{0mm}
\hspace*{0mm}
\psfig{figure=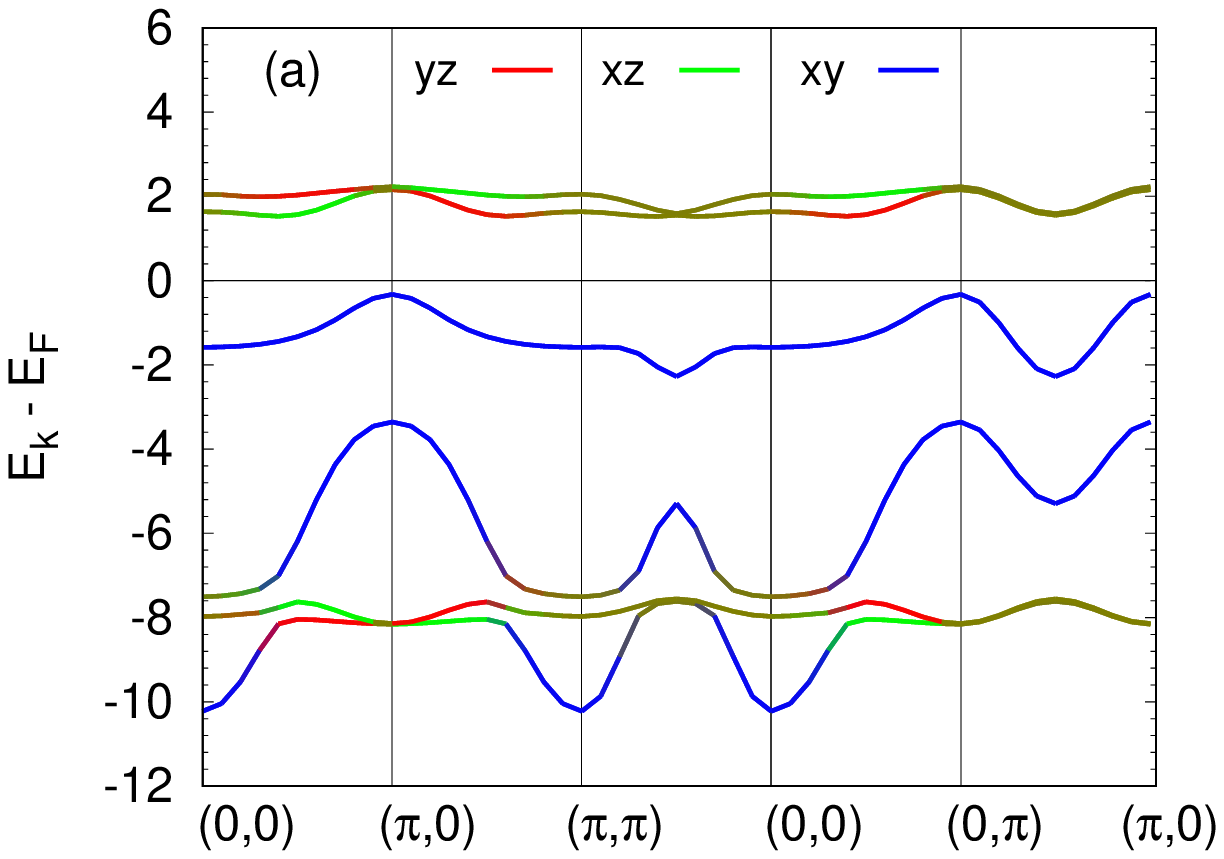,angle=0,width=80mm,angle=0}
\psfig{figure=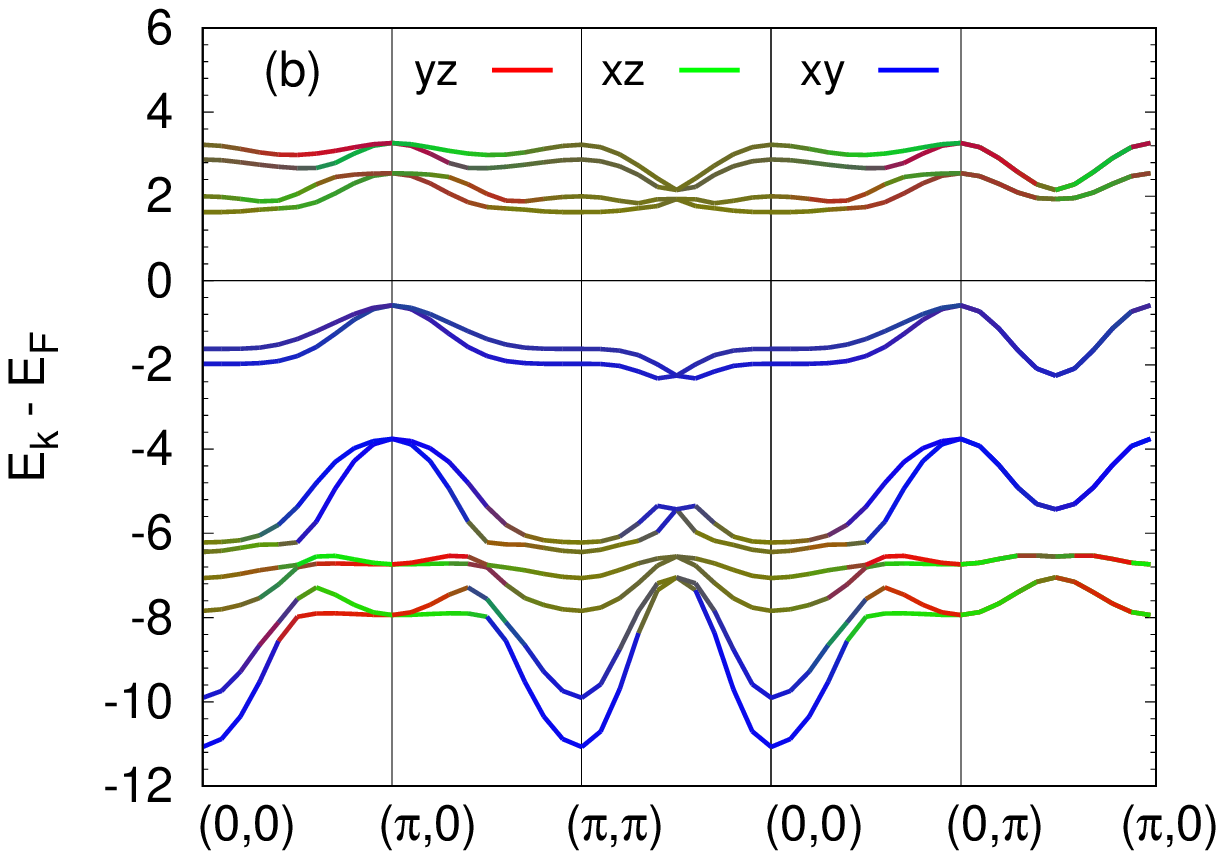,angle=0,width=80mm,angle=0}
\caption{Calculated electronic band structure in the self-consistent AFM state for moderate tetragonal distortion: (a) without and (b) with all off-diagonal spin and charge condensates included, along with octahedral tilting and rotation. Colors indicate dominant orbital weight: red ($yz$), green ($xz$), blue ($xy$). Here $U=8$, $\epsilon_{xy}=-0.8$, and bare SOC = 1.} 
\label{band_all}
\end{figure}

Fig. \ref{band_all} shows the orbital resolved electronic band structure in the self consistent AFM state calculated for the two cases: (a) including only normal condensates, and (b) including all off-diagonal spin and charge condensates along with octahedral rotation and tilting. The band structure shows the narrow AFM sub bands for the magnetically active $yz,xz$ orbitals above and below the Fermi energy due to the dominant exchange field splitting. The relatively smaller splitting between the $xy$ sub bands (both below $E_{\rm F}$) is due to the weaker effect of $yz,xz$ moments through the Hund's coupling. The octahedral tilting and rotation are seen to introduce fine splittings due to the orbital mixing hopping terms. 

\begin{table}[b] \vspace*{-0mm}
\caption{Self consistently determined off-diagonal spin and charge condensates for the three orbital pairs on the two sublattices.} 
\centering 
\begin{tabular}{l r r r r} \\  
\hline\hline 
Orbital pair & $\langle \psi_\mu^\dagger \sigma_x \psi_\nu \rangle$ & $\langle \psi_\mu^\dagger \sigma_y \psi_\nu \rangle$ & $\langle \psi_\mu^\dagger \sigma_z \psi_\nu \rangle$ & $\langle \psi_\mu^\dagger {\bf 1} \psi_\nu \rangle$ \\ [0.5ex]
\hline 
$yz-xz$ (A) & (0.066,0.030) & (0.071,0.025) & (0.018,0.169) & $-$(0.089,0.067) \\ [1ex]
$xz-xy$ (A) & (0.026,0.281) & (0.057,0.126) & (0.079,0.039) & $-$(0.061,0.245) \\ [1ex]
$xy-yz$ (A) & (0.042,0.108) & (0.053,0.333) & (0.081,0.034) & $-$(0.073,0.289) \\[1ex] \hline
$yz-xz$ (B) & $-$(0.071,0.025) & $-$(0.066,0.030) & (0.018,0.169) & $-$(0.089,0.067) \\[1ex]
$xz-xy$ (B) & (0.053,0.333) & (0.042,0.108) & $-$(0.081,0.034) & (0.073,0.289) \\[1ex]
$xy-yz$ (B) & (0.057,0.126) & (0.026,0.281) & $-$(0.079,0.039) & (0.061,0.245) \\[1ex]\hline 
\end{tabular}
\label{table3}
\end{table}

\section{Orbital magnetic moment and SOC renormalization}
The off-diagonal charge condensates $\langle \psi_\mu^\dagger \psi_\nu \rangle$ directly yield the orbital magnetic moments:
\begin{eqnarray}
\langle L_x \rangle &=& \langle \psi_{xz}^\dagger (-i) \psi_{xy} \rangle + \langle \psi_{xy}^\dagger (i) \psi_{xz} \rangle \nonumber \\
&=& -i \langle \psi_{xz}^\dagger \psi_{xy} \rangle + i \langle \psi_{xz}^\dagger \psi_{xy} \rangle^* = 2 {\rm Im} \langle \psi_{xz}^\dagger \psi_{xy} \rangle
\end{eqnarray}
and similarly for the other components. Accordingly, the charge term in Eq. (\ref{h_hf_od}), of which only the anti-symmetric part is non-vanishing (see Appendix), can be represented as a coupling of orbital angular momentum operators to orbital fields:
\begin{eqnarray}
[{\cal H}_{\rm int}^{\rm HF}]_{\rm OOD}^{\rm charge}(i)|_{\rm anti-sym} &=& 
-\frac{U''_{\rm c|a}}{2} \sum_{\mu < \nu} \langle n_{\mu\nu}\rangle^{\rm Im} \left [ \psi_\mu^\dagger (-i) \psi_\nu + {\rm H.c.} \right ] \nonumber \\
&=& -\frac{U''_{\rm c|a}}{4} \left [\langle L_x\rangle L_x + \langle L_y\rangle L_y + \langle L_z\rangle L_z \right ]
\label{orb_int}
\end{eqnarray}
which corresponds to a weak effective isotropic interaction $-(U''_{\rm c|a}/8){\bf L}.{\bf L}$ between orbital moments, and will therefore weakly enhance the $\langle L_\alpha\rangle$ values in the HF calculation.  

Turning now to the spin part of Eq. (\ref{h_hf_od}), the anti-symmetric part (see Appendix) can be represented in terms of the spin-orbital operators:
\begin{eqnarray}
[{\cal H}_{\rm int}^{\rm HF}]_{\rm OOD}^{\rm spin} (i)|_{\rm anti-sym} & = & -(U''_{\rm s|a}/2) \sum_{\mu < \nu} \langle \makebox{\boldmath $\sigma$}_{\mu\nu} \rangle^{\rm Im} . \left [ \psi_\mu ^\dagger (-i \makebox{\boldmath $\sigma$}) \psi_\nu + {\rm H.c.} \right ] \nonumber \\
& = & - \sum_{\alpha=x,y,z} \left [ \lambda_\alpha ^{\rm int} L_\alpha S_\alpha + \sum_{\beta\ne\alpha} \lambda_{\alpha\beta}^{\rm int} L_\alpha S_\beta \right ]
\end{eqnarray}
where the interaction-induced SOC renormalization terms:
\begin{equation}
\lambda_\alpha ^{\rm int} = U''_{\rm s|a} {\rm Im} \langle \psi_\mu^\dagger \sigma_\alpha \psi_\nu \rangle = U''_{\rm s|a} \langle \psi_\mu^\dagger (-i \sigma_\alpha ) \psi_\nu \rangle^{\rm Re} = U''_{\rm s|a} \langle L_\alpha S_\alpha \rangle   
\label{soc_ren} 
\end{equation}
for the orbital pair $\mu,\nu$ corresponding to component $\alpha$. Although the off-diagonal SOC terms $(L_\alpha S_\beta)$ are smaller than the diagonal terms ($\lambda_{\alpha\beta}^{\rm int} < \lambda_\alpha^{\rm int}$), they are still significant. For example, with Im$\langle \psi_{xz}^\dagger \sigma_y \psi_{xy} \rangle =0.126$ from Table II, we obtain $\lambda_{xy}^{\rm int} \approx U'' \times 0.126 \approx 0.5$ on the A sublattice, whereas the bare SOC = 1.0. 

Similarly, for the symmetric part we obtain:
\begin{equation}
[{\cal H}_{\rm int}^{\rm HF}]_{\rm OOD}^{\rm spin} (i)|_{\rm sym} = -(U''_{\rm s|s}/2) \sum_{\mu < \nu} \langle \makebox{\boldmath $\sigma$}_{\mu\nu} \rangle^{\rm Re} . \left [ \psi_\mu ^\dagger  \makebox{\boldmath $\sigma$} \psi_\nu + {\rm H.c.} \right ] 
\end{equation}
representing the coupling of the orbital off-diagonal spin operators to real spin fields involving the enhanced effective interaction $U''_{\rm s|s} = U'' + 3J_{\rm H}/2$. In the limit of bare SOC $\rightarrow 0$, since Im$\langle \psi_\mu^\dagger \psi_\nu \rangle$ and Im$\langle \psi_\mu^\dagger \sigma_\alpha \psi_\nu \rangle$ are identically zero, the above term is the only surviving orbital off-diagonal contribution, and that too only for finite octahedral tilting and rotation which generate orbital mixing. 

\begin{table}
\caption{Self consistently determined renormalized SOC values $\lambda_\alpha = \lambda^{\rm bare}+\lambda_\alpha^{\rm int}$ and the orbital magnetic moments $\langle L_\alpha \rangle$ for $\alpha=x,y,z$ on the two sublattices. Bare SOC strength $\lambda^{\rm bare}=1.0$} 
\centering 
\begin{tabular}{l r r r r r r} \\  
\hline\hline 
$s$ & $\lambda_x$ & $\lambda_y$ & $\lambda_z$ & $\langle L_x \rangle$ & $\langle L_y \rangle$ & $\langle L_z \rangle$ \\ [0.5ex]
\hline 
A & 1.898 & 2.065 & 1.540 & $-$0.490 & $-$0.578 & $-$0.134 \\[1ex]
B & 2.065 & 1.898 & 1.540 & 0.578 & 0.490 & $-$0.134 \\[1ex]
\hline 
\end{tabular}
\label{table2}
\end{table}

We summarize here the results obtained above for moderate tetragonal distortion ($\epsilon_{xy} \sim -1.0$), with all orbital off-diagonal spin and charge condensates included in the self consistent calculation. With nearly half filled $yz,xz$ orbitals and nearly filled $xy$ orbital, the AFM insulating state is characterized by dominantly $yz,xz$ moments lying in the SOC induced easy ($a-b$) plane and aligned along the octahedral tilting induced easy ($b$) axis, with small canting of moments in and about the crystal $c$ axis. The spin cantings become negligible when octahedral tilting and rotation are set to zero. Spin canting in the $c$ direction has been recently observed in resonant elastic X-ray scattering experiments.\cite{porter_PRB_2018} The SOC induced spin-orbital correlations lead to strong orbital moments $\langle L_x \rangle$ and $\langle L_y \rangle$ and strongly anisotropic Coulomb renormalized SOC values ($\lambda_x,\lambda_y > \lambda_z$), as shown in Tables II and III. 

\section{Magnetic reorientation transition}
\begin{figure}
\vspace*{0mm}
\hspace*{0mm}
\psfig{figure=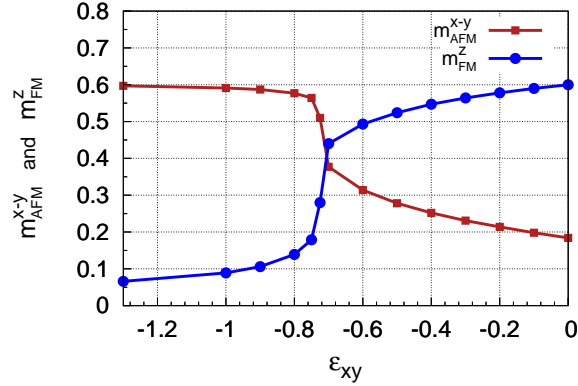,angle=0,width=80mm,angle=0}
\caption{The reorientation transition with decreasing tetragonal distortion, as reflected in the sharp drop in the orbital averaged planar AFM order $m_{\rm AFM}^{x-y}$ and the sharp rise in the FM order $m_{\rm FM}^z$. Here bare SOC = 1.0.} 
\label{reorientation}
\end{figure}

With decreasing tetragonal distortion, we find a sharp magnetic reorientation transition from the dominantly $a-b$ plane AFM order to a dominantly $c$ axis FM order, as shown in Fig. (\ref{reorientation}). The two orbital averaged magnetic orders shown in this plot are defined as:
\begin{eqnarray}
m_{\rm AFM}^{x-y} &=& (1/3)\sum_\mu \left [ \left (\frac{m_\mu^x (A) - m_\mu^x (B)}{2}\right )^2 +  \left (\frac{m_\mu^y (A) - m_\mu^y (B)}{2}\right )^2 \right ] ^{1/2} \nonumber \\
m_{\rm FM}^z &=& (1/3)\sum_\mu m_\mu^z
\end{eqnarray}
The planar AFM order decreases sharply across the transition, while the FM ($z$) order (which is same for both sublattices) increases sharply. The electronic state remains insulating throughout the range of $\epsilon_{xy}$ shown, with filling $n=4$. AFM correlations are seen to persist after the transition to the FM ($z$) order.  

\begin{figure}
\vspace*{0mm}
\hspace*{0mm}
\psfig{figure=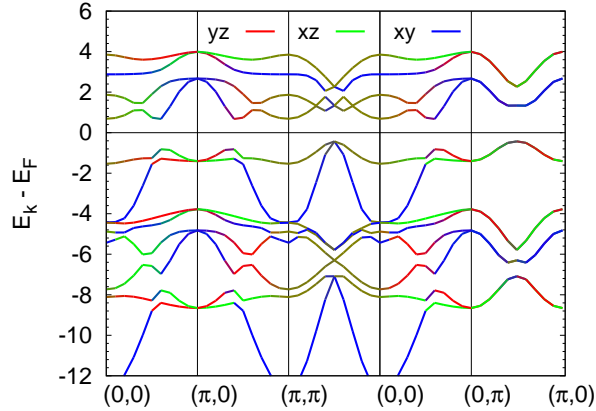,angle=0,width=80mm,angle=0}
\caption{Orbital resolved electronic band structure for the FM ($z$) order ($n=4$), obtained for reduced tetragonal distortion, with no octahedral tilting or rotation. Here bare SOC = 0.5 and $\epsilon_{xy}=-0.5$.} 
\label{fig6}
\end{figure}

The reorientation transition is even stronger for bare SOC = 0.5 which corresponds to the realistic value of 100 meV. Results for the FM ($z$) order obtained for $\epsilon_{xy}=-0.5$ with no octahedral tilting or rotation are particularly interesting, with identical magnetization ($m_\mu ^z=0.65$) and density ($n_\mu=4/3$) for all three orbitals, and very small planar components $m_\mu^{x,y}$. The renormalized SOC and orbital moment values obtained are: $\lambda_{x,y,z}= (0.78,0.78,1.28)$ and $\langle L_{x,y,z}\rangle=(\mp 0.26,\mp 0.26,-0.48)$ on A/B sublattice. The electronic band structure in the self consistent state is shown in Fig. \ref{fig6} for this case. We find that the indirect band gap between valence band top at ($\pi/2,\pi/2$) and conduction band bottom at ($\pi,\pi$), ($0,0$) is reduced to nearly zero for slightly enhanced $yz,xz$ NN hopping term corresponding to no octahedral tilting. 

Fig. \ref{fig6} also shows the small orbital gap near the Fermi energy highlighting the orbital physics. Band splittings near $(\pi/2,0)$, $(\pi,\pi/2)$, and $(0,\pi/2)$ arise from the orbital moment interaction term (Eq. \ref{orb_int}). Finite $\langle L_x \rangle$ and $\langle L_y \rangle$ generate orbital fields which couple to the orbital angular momentum operators involving mixings between $xy$ and $yz,xz$ orbitals. The consequent orbital field induced splitting is analogous to the usual exchange field splitting of spin sub bands. The small orbital gap vividly illustrates the crucial role of the orbital off-diagonal charge condensates in the insulating behaviour. With increasing $\epsilon_{xy}$ pushing up the $xy$ bands, the upper $xy$ sub-band is now seen to be straddling the orbital gap, reflecting an important interplay between orbital physics and decreasing tetragonal distortion. The orbital gap is maintained even as the $xy$ spectral weight is transferred across the Fermi energy. 
 
We also find a robust FM metallic phase for electron filling $n\gtrsim 4$. Results of the self consistent cacluation obtained for the same set of parameters as above (bare SOC = 0.5 and $\epsilon_{xy}=-0.5$) are shown in Table IV. The FM metallic phase is characterized by identically vanishing planar magnetization components. All orbital off-diagonal condensates except for the SOC renormalization terms $\langle L_\alpha S_\alpha \rangle$ are also identically zero. Driven by switching of the dominant role from $yz,xz$ orbitals (AFM interaction) to the $xy$ orbital (FM interaction), the magnetic reorientation transition with decreasing tetragonal distortion as discussed above provides a unified understanding of the planar AFM order as well as the low-temperature FM metallic phase found in $\rm Ca_2 Ru O_4$ under high pressure\cite{nakamura_PRB_2002} and also in $\rm Ca_{2-x} Sr_x Ru O_4$ for $x \sim 0.5$ in neutron and DFT studies.\cite{nakatsuji_PRB_2000,friedt_PRB_2001,fang_PRB_2001,fang_PRB_2004}

\begin{table}[b]
\caption{Self consistently determined magnetization and density values, along with renormalized SOC and orbital magnetic moment values in the FM metallic phase, with bare SOC = 0.5, $\epsilon_{xy}=-0.5$, and no octahedral rotation and tilting.} 
\centering 
\begin{tabular}{c c c c c} \\  
\hline\hline 
$\mu$ & $m_\mu^x$ & $m_\mu^y$ & $m_\mu^z$ & $n_\mu$ \\ [0.5ex]
\hline 
$yz$ & 0 & 0 & 0.57 & 1.42 \\[1ex]
$xz$ & 0 & 0 & 0.57 & 1.42 \\[1ex]
$xy$ & 0 & 0 & 0.66 & 1.33 \\[1ex] \hline 
\end{tabular} \hspace{10mm} 
\begin{tabular}{c c c c c c} \\  
\hline\hline 
$\lambda_x$ & $\lambda_y$ & $\lambda_z$ & $\langle L_x \rangle$ & $\langle L_y \rangle$ & $\langle L_z \rangle$ \\ [0.5ex]
\hline 
0.73 & 0.73 & 1.55 & $0$ & $0$ & $-0.65$ \\[1ex]
\hline 
\end{tabular}
\label{table4}
\end{table}

\begin{figure}
\vspace*{0mm}
\hspace*{0mm}
\psfig{figure=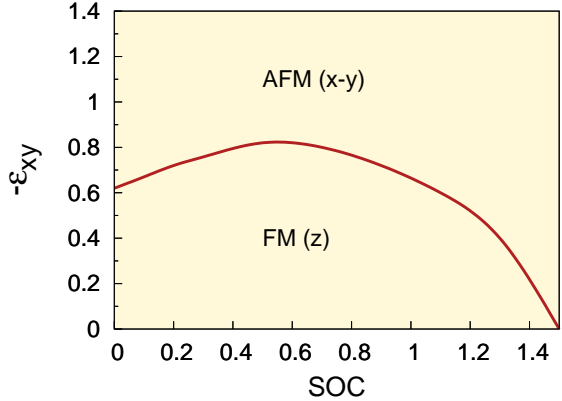,angle=0,width=80mm,angle=0} 
\caption{The magnetic phase boundary between the two regions with dominantly planar AFM and FM ($z$) orders.} 
\label{boundary}
\end{figure}

For higher values of bare SOC, the planar AFM order is stable even for reduced tetragonal distortion, which is expected from the SOC induced easy $a-b$ plane anisotropy. However, in the weak SOC regime (bare SOC $\lesssim 0.5$), the FM ($z$) order is stabilized with increasing SOC, as seen in Fig. \ref{boundary}, which shows the phase boundary between the two magnetic orders. The two axes here represent increasing bare SOC and tetragonal distortion. For realistic value of bare SOC = 0.5 and slightly above the magnetic phase boundary ($\epsilon_{xy}=-1.0$), we also find a stable AFM metallic state for $n\lesssim 4$, suggesting persistence of AFM correlations even if long range AFM order is destroyed by quantum spin fluctuations as in cuprate antiferromagnets. This is in agreement with the antiferromagnetically correlated metallic state reported for $\rm Ca_{2-x} Sr_x Ru O_4$ in the range $0.2<x<0.5$. 

\begin{figure}
\vspace*{0mm}
\hspace*{0mm}
\psfig{figure=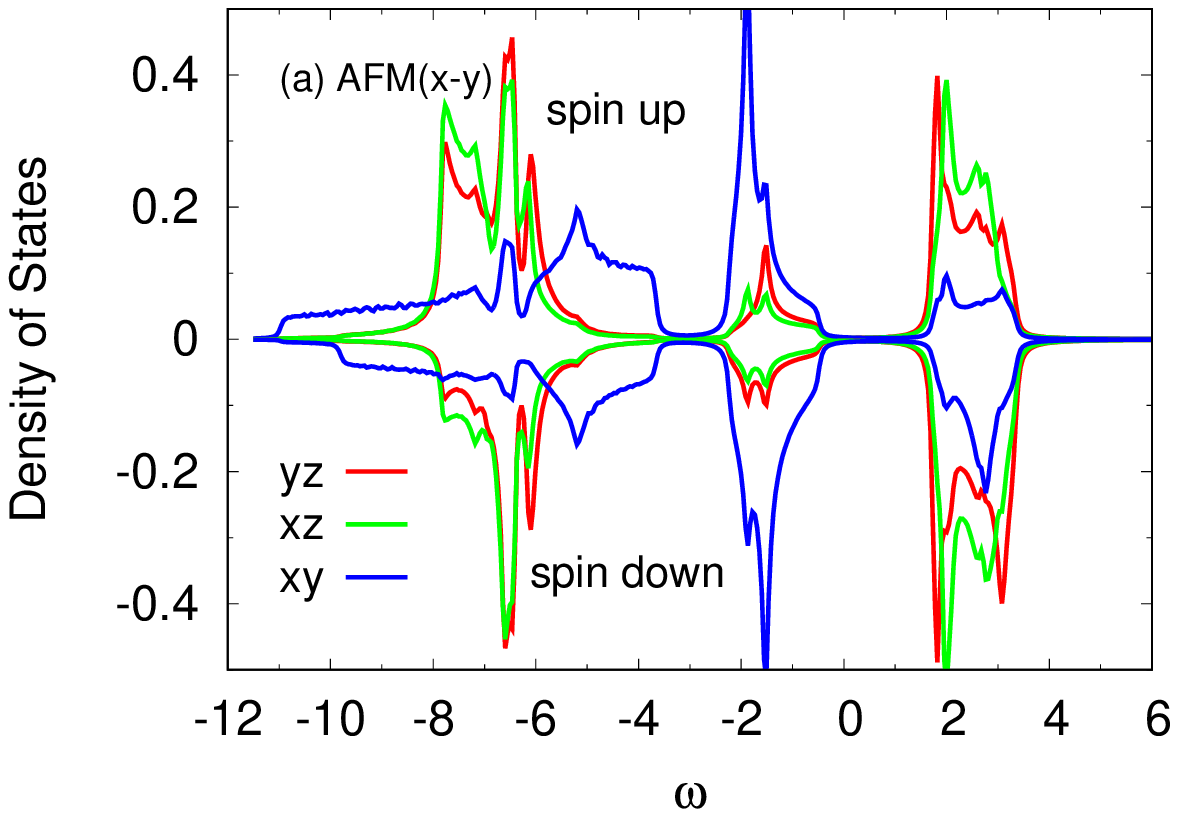,angle=0,width=80mm,angle=0} 
\psfig{figure=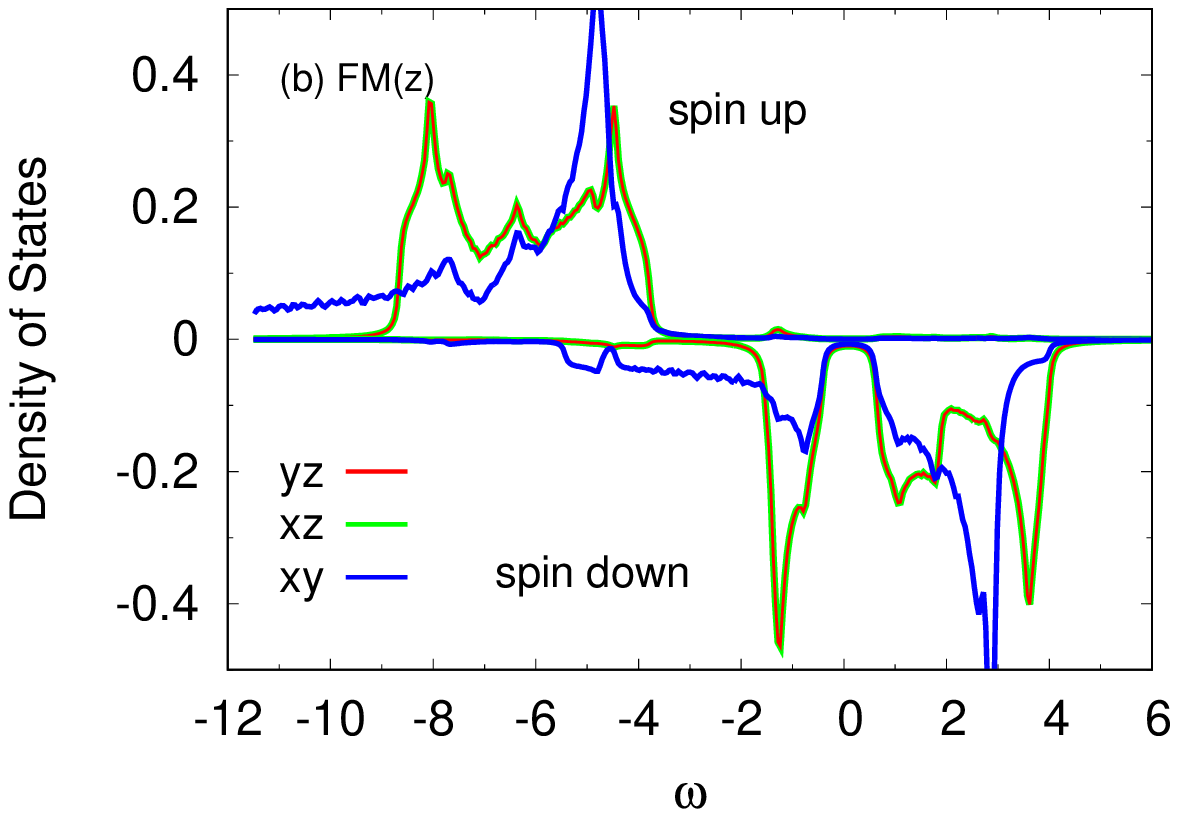,angle=0,width=80mm,angle=0} 
\caption{The spin-resolved electronic density of states for the (a) planar AFM order and (b) FM ($z$) order, with same parameters as in Fig. 3(b) and Fig. 5.} 
\label{dos}
\end{figure}

Spin resolved electronic density of states (DOS) is shown in Fig. \ref{dos} for (a) planar AFM and (b) FM ($z$) order, with same parameters as in Figs. 3(b) and 5. For FM ($z$) order, Fig. \ref{dos}(b) shows that states near the Fermi energy are purely minority (down) spin states, highlighting the orbital character of the small gap as discussed for Fig. 5. Also, spin down spectral weight for the $xy$ orbital is transferred above the Fermi energy, whereas for $yz,xz$ orbitals it is transferred below, reversing the dominant orbital weight in the sub band just below the Fermi energy from $xy$ (planar AFM order) to $yz,xz$ (FM order). 

The importance of the orbital off-diagonal spin and charge condensates in determining the self consistent magnetic order is illustrated by the strongly anisotropic SOC renormalization and strong orbital magnetic moments, which are both magnetic order dependent. We also note here that without the off-diagonal condensates included in the self consistent calculation, the planar AFM order is obtained even for reduced tetragonal distortion (down to $\epsilon_{xy} = -0.3$). The off-diagonal condensates are therefore responsible for the reorientation transition from planar AFM order to FM ($z$) order. 

The reduced tetragonal distortion induced reorientation transition as found here provides a microscopic understanding of the pressure-induced stabilization of FM order in $\rm Ca_2RuO_4$ and the chemical substitution induced stabilization of FM correlations in the isoelectronic series $\rm Ca_{2-x}Sr_x RuO_4$. Another candidate for the theory presented here is the $\rm Ca_2RuO_4$ thin film where the tetragonal distortion and octahedral tilting are tuned by the film thickness, as found in the recently synthesized nanofilm single crystal,\cite{nobukane_SREP_2020} which shows robust FM correlations and significantly higher Curie temperature ($T_{\rm C}=180$ K) due to the suppression of lattice distortion. Other possible candidates could be ruthenate heterostructures where lattice distortions are tuned by synthesizing layered superlattices, as in the recently studied bilayer iridate heterostructure.\cite{meyers_SREP_2019}

\section{Coupled spin-orbital fluctuations}
\begin{figure}
\vspace*{0mm}
\hspace*{0mm}
\psfig{figure=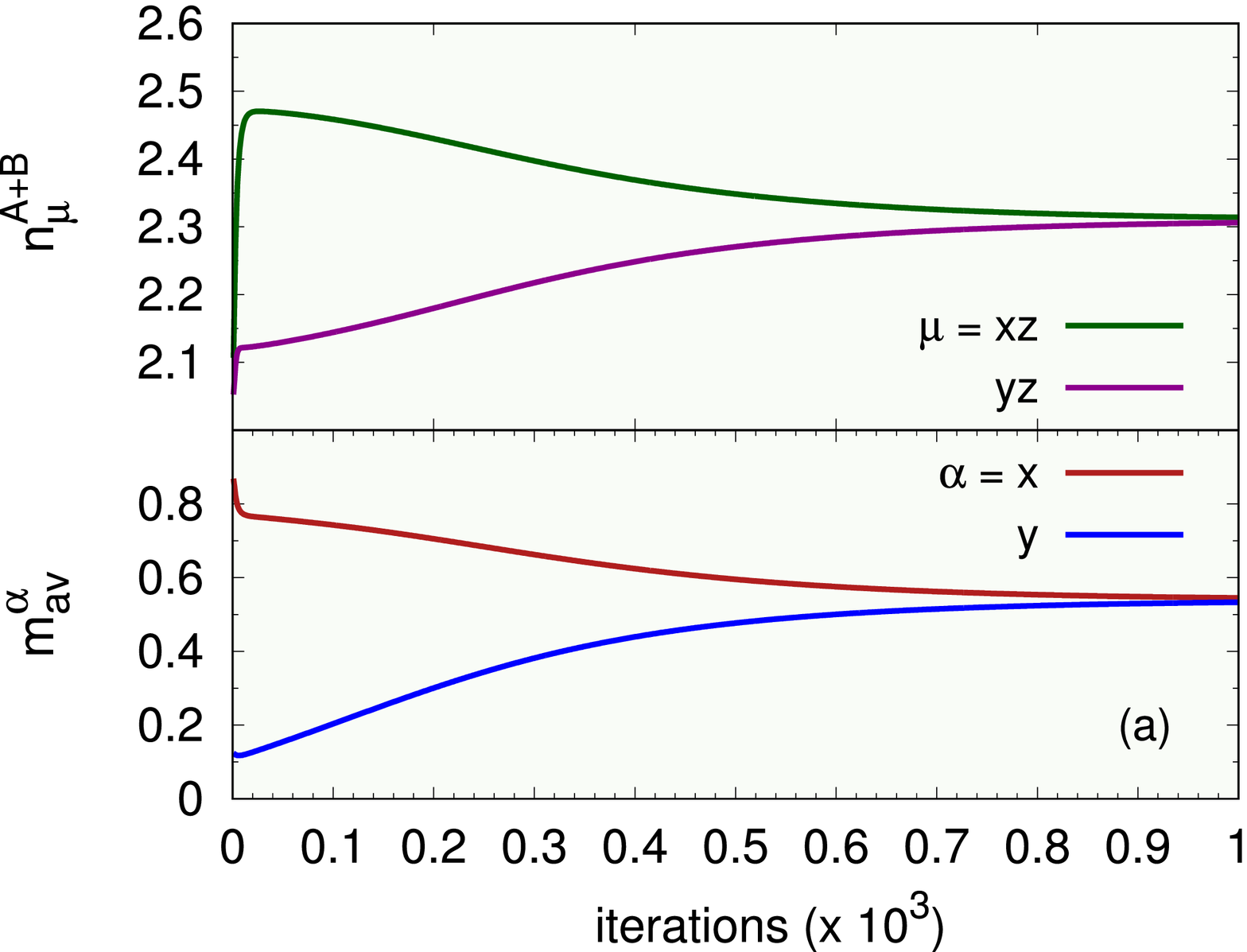,angle=0,width=80mm,angle=0}
\psfig{figure=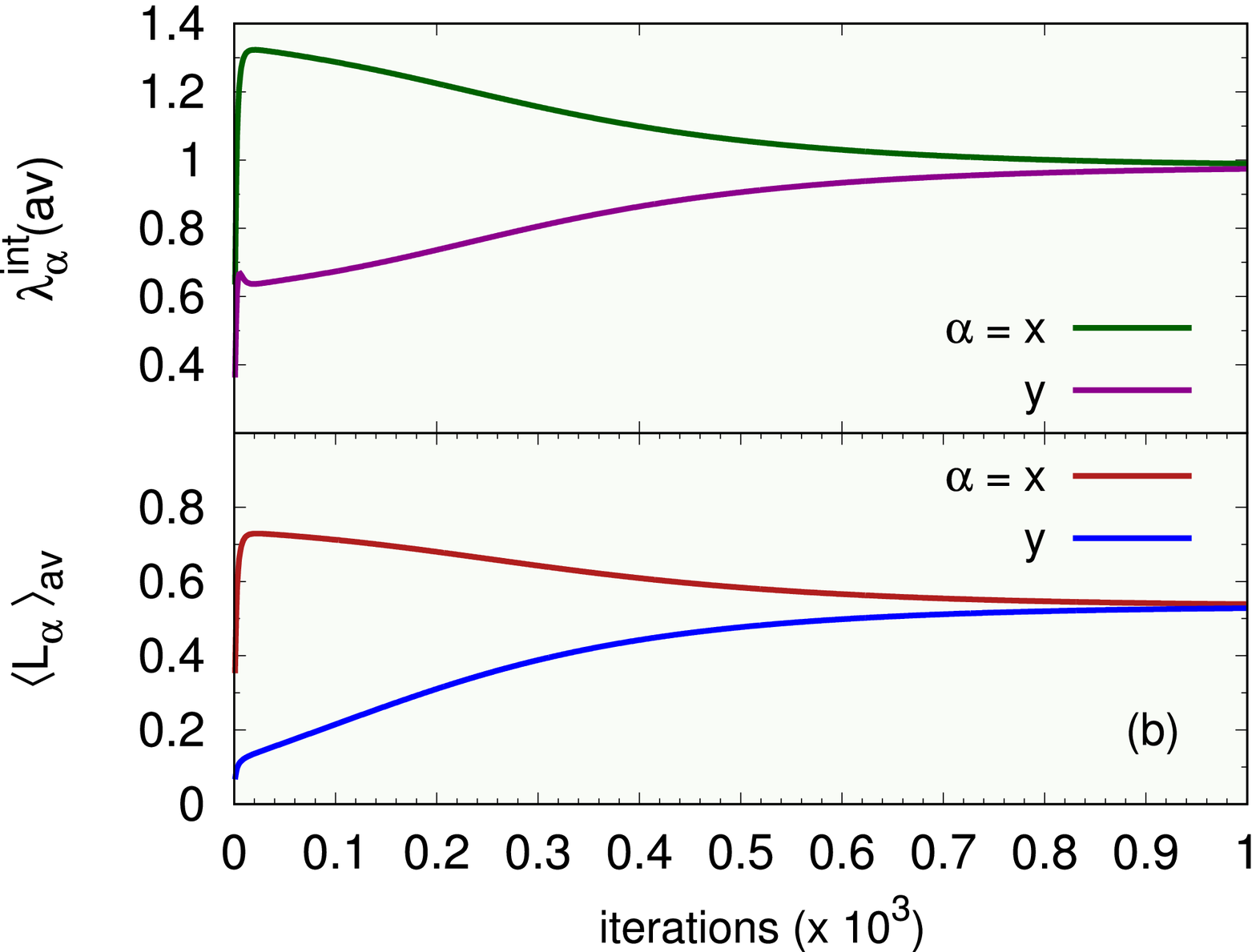,angle=0,width=80mm,angle=0}
\caption{Variation of the (a) $yz,xz$ orbital densities (upper panel) and $x,y$ components of the sublattice magnetization (lower panel), (b) $x,y$ components of the interaction induced SOC renormalizations (upper panel) and orbital magnetic moments (lower panel), with iteration in the self consistency process for the planar AFM order. Here bare SOC $=1.0$ and $\epsilon_{xy}=-0.8$.} 
\label{iter}
\end{figure}

Spin orientation in the AFM state affects orbital densities due to strong spin-orbital coupling in $\rm Ca_2RuO_4$. Fig. \ref{iter}(a) shows the variation of $yz,xz$ orbital densities (summed over both sublattices) with iterations in the self-consistency process, starting with spins oriented towards the $x$ direction. Also shown are the sublattice magnetization components $m_{\rm av}^x$ and $m_{\rm av}^y$ averaged for $yz,xz$ orbitals. Initially, we find that $n_{xz} > n_{yz}$, whereas the two densities converge as the spin orientation approaches the self-consistent easy-axis ($\phi=\pi/4$) direction. This implies that the planar Goldstone mode, corresponding to rigid spin rotation away from the easy axis towards $x$ ($y$) axis, will be associated with ferro orbital fluctuation due to density transfer between orbitals. 
In contrast, the out-of-phase (zone boundary) fluctuation mode, with spin twistings towards $x$ ($y$) and $-y$ ($-x$) directions on A and B sublattices, respectively, will be associated with antiferro orbital fluctuation with opposite sign of $n_{xz}-n_{yz}$ on the two sublattices. The physical quantities related to orbital off-diagonal condensates also show [Fig. \ref{iter}(b)] strong dependence on the spin orientation.  

Quite generally, since the self consistent determination of magnetic order requires all spin and charge condensates to be included, investigation of the fluctuation propagator must therefore necessarily involve the generalized spin ($\psi_\mu ^\dagger \makebox{\boldmath $\sigma$} \psi_\nu$) and charge ($\psi_\mu ^\dagger \psi_\nu$) operators including both orbital diagonal and off-diagonal parts. This requires consideration of the generalized time-ordered fluctuation propagator:
\begin{equation}
[\chi({\bf q},\omega)] = \int dt \sum_i e^{i\omega(t-t')} 
e^{-i{\bf q}.({\bf r}_i -{\bf r}_j)} 
\times \langle \Psi_0 | T [O_{\mu\nu}^\alpha (i,t) O_{\mu'\nu'}^{\alpha'} (j,t')] |\Psi_0 \rangle 
\end{equation}
in the self-consistent AFM ground state $|\Psi_0 \rangle$, where the generalized spin-charge operators at lattice sites $i,j$ are defined as $O_{\mu\nu}^\alpha = \psi_\mu ^\dagger \sigma^\alpha \psi_\nu$, which include both the orbital diagonal ($\mu=\nu$) and off-diagonal ($\mu\ne\nu$) cases, and the spin ($\alpha=x,y,z$) and charge ($\alpha=c$) operators, with $\sigma^\alpha$ defined as Pauli matrices for $\alpha=x,y,z$ and unit matrix for $\alpha=c$. 

Investigation of the generalized fluctuation propagator can reveal if the planar Goldstone mode acquires a finite mass due to the coupled spin-orbital fluctuations, as reflected in the ferro and antiferro orbital fluctuations associated with in-phase and out-of-phase spin twisting modes. The coupling between spin and orbital fluctuations clearly highlights the strong deviation from conventional Heisenberg behaviour in effective spin models, as discussed recently to account for the magnetic excitation measurements in INS experiments.\cite{jain_NATPHY_2017}

\section{Conclusions}

Including the orbital off-diagonal spin and charge condensates in the self consistent determination of magnetic order illustrates the rich interplay between the different physical elements in the $\rm 4d^4$ compound $\rm Ca_2 Ru O_4$. These include SOC induced easy-plane anisotropy, octahedral tilting induced easy-axis anisotropy, spin-orbital coupling induced orbital magnetic moments, Coulomb interaction induced anisotropic SOC renormalization, decreasing tetragonal distortion induced magnetic reorientation transition from planar AFM order to FM ($z$) order, and orbital moment interaction induced orbital gap. Stable FM and AFM metallic states were also obtained near the magnetic phase boundary separating the two magnetic orders. Since the orbital off-diagonal condensates contribute on the same footing as the normal condensates, the coupled spin-orbital fluctuations must be investigated within a unified formalism involving the generalized spin and charge operators including orbital off-diagonal terms.

\newpage 
\appendix*

\section{Orbital off-diagonal condensates in the HF approximation}

The additional contributions in the HF approximation arising from the orbital off-diagonal spin and charge condensates are given below. For the density, Hund's coupling, and pair hopping interaction terms in Eq. \ref{h_int}, we obtain (for site $i$):
\begin{eqnarray}
U'' \sum_{\mu < \nu} n_\mu n_\nu & \rightarrow & 
-\frac{U''}{2} \sum_{\mu < \nu} \left [ n_{\mu\nu} \langle n_{\nu\mu} \rangle + \makebox{\boldmath $\sigma$}_{\mu\nu}.\langle \makebox{\boldmath $\sigma$}_{\nu\mu} \rangle \right ] + {\rm H.c.} \nonumber \\
-2J_{\rm H} \sum_{\mu < \nu} {\bf S}_\mu . {\bf S}_\nu 
& \rightarrow & \frac{J_{\rm H}}{4} \sum_{\mu < \nu} 
\left [3\, n_{\mu\nu} \langle n_{\nu\mu} \rangle - \makebox{\boldmath $\sigma$}_{\mu\nu}.\langle \makebox{\boldmath $\sigma$}_{\nu\mu} \rangle \right ] + {\rm H.c.} \nonumber \\
J_{\rm P} \sum_{\mu \ne \nu} a_{\mu\uparrow}^{\dagger} a_{\mu\downarrow}^{\dagger} a_{\nu\downarrow} a_{\nu\uparrow} & \rightarrow &  
\frac{J_{\rm P}}{2} \sum_{\mu < \nu} \left [n_{\mu\nu} \langle n_{\mu\nu} \rangle - \makebox{\boldmath $\sigma$}_{\mu\nu}.\langle \makebox{\boldmath $\sigma$}_{\mu\nu} \rangle \right ] + {\rm H.c.} 
\end{eqnarray}
in terms of the orbital off-diagonal spin ($\makebox{\boldmath $\sigma$}_{\mu\nu}=\psi_\mu^\dagger \makebox{\boldmath $\sigma$} \psi_\nu$) and charge ($n_{\mu\nu} = \psi_\mu^\dagger {\bf 1} \psi_\nu$) operators. The orbital off-diagonal condensates are finite due to the SOC-induced spin-orbital correlations. These additional terms in the HF theory explicitly preserve the SU(2) spin rotation symmetry of the various Coulomb interaction terms. 

Collecting all the spin and charge terms together, we obtain the orbital off-diagonal (OOD) contributions of the Coulomb interaction terms:
\begin{eqnarray}
[{\cal H}_{\rm int}^{\rm HF}]_{\rm OOD} &=& \sum_{\mu < \nu} \left [ \left (-\frac{U''}{2} + \frac{3J_{\rm H}}{4} \right ) 
n_{\mu\nu} \langle n_{\nu\mu} \rangle + \left (\frac{J_{\rm P}}{2}\right ) n_{\mu\nu} \langle n_{\mu\nu} \rangle \right . \nonumber \\
& & - \left . \left (\frac{U''}{2} + \frac{J_{\rm H}}{4} \right )\makebox{\boldmath $\sigma$}_{\mu\nu}.\langle \makebox{\boldmath $\sigma$}_{\nu\mu} \rangle - \left (\frac{J_{\rm P}}{2} \right )\makebox{\boldmath $\sigma$}_{\mu\nu}.\langle \makebox{\boldmath $\sigma$}_{\mu\nu} \rangle \right ] + {\rm H.c.} 
\end{eqnarray}
Separating the condensates $\langle n_{\mu\nu} \rangle = \langle n_{\mu\nu} \rangle^{\rm Re} + i \langle n_{\mu\nu} \rangle^{\rm Im}$ into real and imaginary parts in order to simplify using $\langle n_{\nu\mu}\rangle=\langle n_{\mu\nu}\rangle^*$, and similarly for $\langle \makebox{\boldmath $\sigma$}_{\mu\nu} \rangle$, allows for organizing the OOD charge and spin  contributions into orbital symmetric and anti-symmetric parts:
\begin{eqnarray}
[{\cal H}_{\rm int}^{\rm HF}]_{\rm OOD} &=& -\frac{U''_{\rm c|s}}{2} \sum_{\mu < \nu} \langle n_{\mu\nu} \rangle^{\rm Re} \left [ n_{\mu\nu} + {\rm H.c.} \right ] - \frac{U''_{\rm c|a}}{2} \sum_{\mu < \nu} \langle n_{\mu\nu} \rangle^{\rm Im} \left [ -i n_{\mu\nu} + {\rm H.c.} \right ] \nonumber \\
& - & \frac{U''_{\rm s|s}}{2} \sum_{\mu < \nu} 
\langle \makebox{\boldmath $\sigma$}_{\mu\nu} \rangle^{\rm Re} . 
\left [ \makebox{\boldmath $\sigma$}_{\mu\nu} + {\rm H.c.} \right ] - \frac{U''_{\rm s|a}}{2} \sum_{\mu < \nu} 
\langle \makebox{\boldmath $\sigma$}_{\mu\nu} \rangle^{\rm Im} . 
\left [ -i \makebox{\boldmath $\sigma$}_{\mu\nu} + {\rm H.c.} \right ]
\end{eqnarray}
where the effective interaction terms above are obtained as: 
\begin{eqnarray}
U''_{\rm c|a} &=& U''_{\rm s|a} = U'' - J_{\rm H}/2 = U - 3J_{\rm H} \nonumber \\
U''_{\rm s|s} &=& U'' + 3J_{\rm H}/2 = U - J_{\rm H} \nonumber \\
U''_{\rm c|s} &=& U''-5J_{\rm H}/2 = U-5J_{\rm H} 
\end{eqnarray}
using $J_{\rm P}=J_{\rm H}$. While the effective interaction $U''_{\rm s|s}$ (spin term, symmetric part) is enhanced relative to $U''$, the corresponding charge term interaction $U''_{\rm c|s}$ vanishes for $J_{\rm H}=U/5$.

\end{document}